\documentclass[sigconf]{acmart}

\usepackage{balance}
\usepackage{enumitem}
\setlist{leftmargin=*}

\AtBeginDocument{%
  }

\copyrightyear{2025}
\acmYear{2025}
\setcopyright{rightsretained}
\acmConference[OZCHI '25]{37th Australian Conference on Human-Computer Interaction (HCI)}{November 29-December 03, 2025}{Sydney, Australia}
\acmBooktitle{37th Australian Conference on Human-Computer Interaction (HCI) (OZCHI '25), November 29-December 03, 2025, Sydney, Australia}
\acmPrice{}
\acmDOI{10.1145/3764687.3764719}
\acmISBN{979-8-4007-2016-1/25/11}

\begin{document}

\setlength{\textfloatsep}{3pt plus 2pt minus 2pt}   
\setlength{\intextsep}{3pt plus 2pt minus 2pt}  
\setlength{\floatsep}{3pt plus 2pt minus 2pt} 

\setlength{\abovecaptionskip}{3pt plus 1pt minus 1pt}
\setlength{\belowcaptionskip}{2pt}

\AtBeginEnvironment{quote}{%
  \setlength{\topsep}{0.75pt}%
  \setlength{\partopsep}{0pt}%
  \fontsize{8.25pt}{9.5pt}\selectfont 
}

\title{Imagining Design Workflows in Agentic AI Futures}

\orcid{0002-0953-5306}
\author{Samangi Wadinambiarachchi}
\affiliation{%
  \institution{The University of Melbourne}
  \city{Melbourne}
  \country{Australia}}
\email{samangi.w@unimelb.edu.au}

\author{Jenny Waycott}
\affiliation{%
  \institution{The University of Melbourne}
  \city{Melbourne}
  \country{Australia}}
\email{jwaycott@unimelb.edu.au}

\author{Yvonne Rogers}
\affiliation{%
  \institution{University College London}
  \city{London}
  \country{United Kingdom}}
\email{y.rogers@ucl.ac.uk}

\author{Greg Wadley}
\affiliation{%
  \institution{The University of Melbourne}
  \city{Melbourne}
  \country{Australia}}
\email{greg.wadley@unimelb.edu.au}

\renewcommand{\shortauthors}{Wadinambiarachchi et al.}


\begin{abstract}

As designers become familiar with \textit{generative AI}, a new concept is emerging: \textit{agentic AI}. While generative AI produces output in response to prompts, agentic AI systems promise to perform mundane tasks autonomously, potentially freeing designers to focus on what they love: being creative. But how do designers feel about integrating agentic AI systems into their workflows? Through design fiction, we investigated how designers want to interact with a collaborative agentic AI platform. Ten professional designers imagined and discussed collaborating with an AI agent to organise inspiration sources and ideate. Our findings highlight the roles AI agents can play in supporting designers, the division of authority between humans and AI, and how designers’ intent can be explained to AI agents beyond prompts. We synthesise our findings into a conceptual framework that identifies authority distribution among humans and AI agents and discuss directions for utilising AI agents in future design workflows.

\end{abstract}




\maketitle
\section{Introduction}

Imagine a group of designers gathered around a table with their client. They aim to understand the client’s preference for the visual look and feel of a digital product and to identify which visual concept they should explore for their newest app. The User Experience (UX) lead shows a wire-flow while the rest of the team suggests different visual concepts and ideas that might suit the app. In one corner of the table, a User Interface (UI) designer struggles to find a saved picture they want to show the others because they believe it is the best inspirational stimulus. While listening to the discussion, the client sitting on the other side of the table suggests some specific colours different to those suggested by the designers. Does this scenario sound familiar? This is common in collaborative design settings, where finding inspiration sources for a task and agreeing upon concepts becomes challenging~\cite{Gonalves2016}.

These problems in the early stages of design are caused by difficulty in articulating abstract concepts with appropriate visual aids~\cite{Davis1999}, difficulty in identifying what is inspiring, not having a proper way to manage ideas and inspiration archives~\cite{Keller2006, inie2018designing}, getting fixated on the external stimuli designers or clients are exposed to~\cite{Wadinambiarachchi2024,Youmans2014a} and time pressure~\cite{Cajander2022}.

Recently, generative AI (GenAI) technologies have been rapidly integrated into design software to help designers overcome difficulties and make design workflows more efficient~\cite{Chung2022, Joon2021, Ning2023}. GenAI tools can support different tasks in the design process, such as generating images, brainstorming, writing micro-copy, creating design variations, and assisting designers with some of the challenges mentioned earlier.

Concurrently, a new paradigm called agentic AI is emerging, which refers to "autonomous systems designed to pursue complex goals with minimal human intervention"~\cite[p.~1]{acharya2025agentic}. Unlike GenAI systems, which enable users to generate content such as images, text, audio, or video based on trained data by using natural language prompts~\cite{koch2020imagesense}, agentic AI systems are goal-oriented and perform activities with limited direct human supervision~\cite{acharya2025agentic, shavit2023practices} enabling designers to co-create with AI, allowing users to refine their work with personalised suggestions, transforming  design work with in-context learning rather than limiting to trained data sets, and promoting collaborations~\cite{Chung2022, Joon2021}.

The field of AI-powered tools is becoming a central focus for the Human-Computer Interaction (HCI) community. Recent work on AI tools reports both positive and negative effects. Limitations include overreliance on textual prompts as the primary input mode~\cite{Peng2024,Takaffoli2024}, the restriction that interfaces are developed primarily for single users, and the limited use of AI in a collaborative setting~\cite{johnson2025exploring}. HCI lacks empirical evidence associating the user perspective with designing these emerging tools, providing an opportunity for human-centred AI. However, the field is nascent; research is needed to understand the potential impact of agentic AI in design workflows, designers’ views on what they need, and how AI should be best designed. Therefore, this study aimed \textbf{to understand how designers want AI agents’ support to enhance the design process, while identifying opportunities and challenges for designing future agentic AI-powered creativity support tools} (A-AICSTs). We invited designers to imagine how an AI agent(s) could support them to access inspiration sources and perform ideation \textit{to expand their conceptual space (abstract mental frameworks enabling novel and valuable ideas)}~\cite{Boden2009, Chen2023} in a design project.

We conducted a design fiction study with 10 professional designers, incorporating a novel interactive story card method, \textbf{\textit{Flip-flap story cards}}, to elicit participant responses through storytelling and critique~\cite{Halskov2006, Rogerson2022}. To derive insights, we employed Braun and Clarke’s six-phase reflexive thematic analysis approach~\cite{Braun2022, Terry2021}, seeking to address the following research questions:

\begin{itemize}
    \item RQ 1: How do designers envision the presence of AI agents in their design workflows?
    \item RQ 2: How should design tasks and authority be distributed between human designers and AI?
    \item RQ 3: What are the design opportunities for future agentic AI systems for designers?
\end{itemize}

Our contribution to HCI is threefold. First, we make an empirical contribution of insights into the users’ point of view on roles for AI agents, the authority and boundaries between human and AI, and how the communication can be envisioned beyond prompts in agentic AI-powered creativity support tools. Second, we introduce a new method for data collection: Flip-Flap story cards, used to elicit participant responses through storytelling and critique. These cards were used in our design fiction study to collect participants’ insights on conceptual solutions for design problems. Third, we synthesise our findings into a conceptual framework for authority distribution among humans and AI when designing future A-AICSTs.

\section{Related Work}

In this section, we describe inspiration in the design process, outline the current state of GenAI in design workflows, and introduce the concept of agentic AI, while highlighting the opportunities for exploring integrating AI to access inspirational stimuli and expand conceptual space, which motivated our design fiction study.

\subsection{Inspiration in the Design Process}

\begin{figure}[h!]
    \centering
    \includegraphics[width=0.75\linewidth]{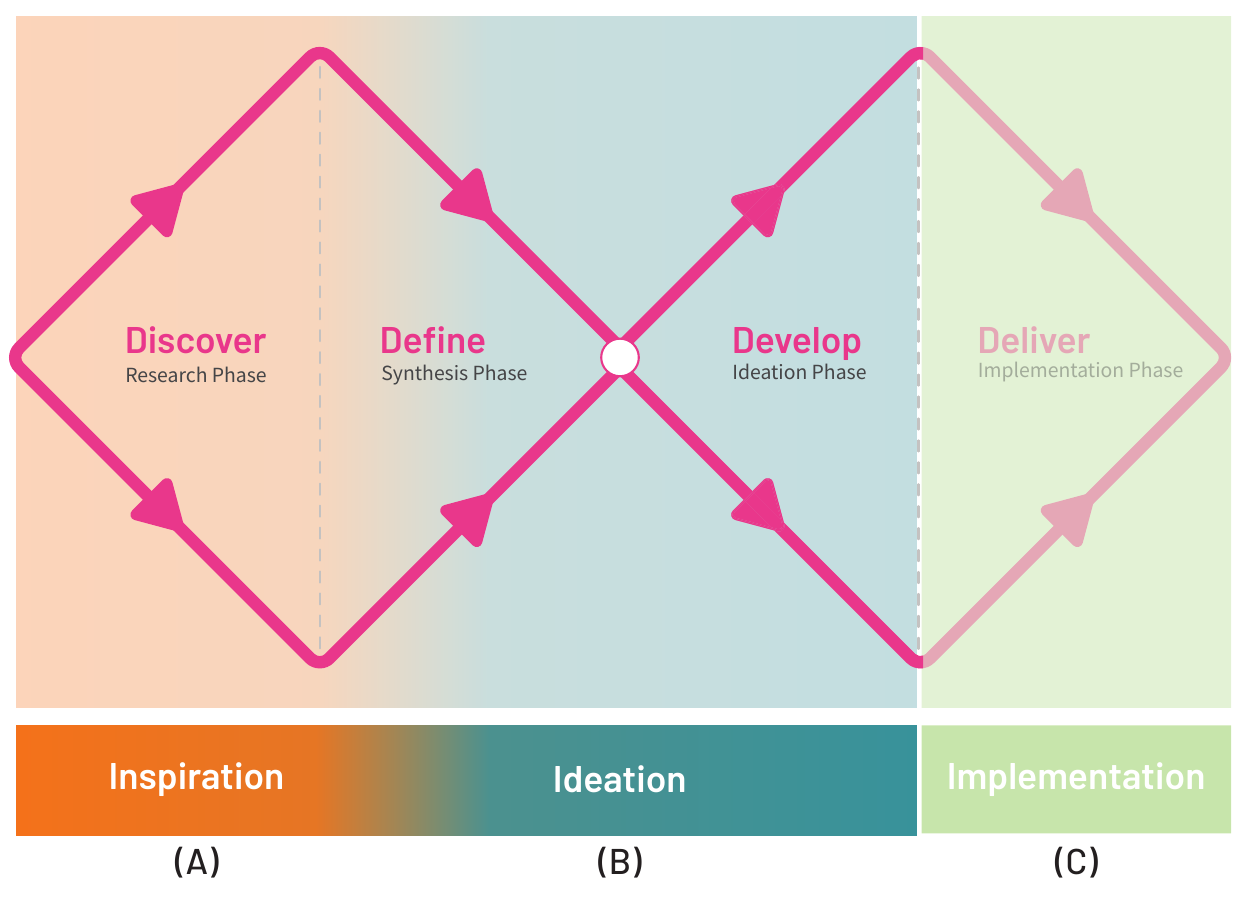}
    \caption{Inspiration in design thinking process, illustrated by adopting IDEO’s three core activities of design thinking~\cite{ideo2015field} and the Double Diamond framework~\cite{DesignCouncil2024}}.
    \label{fig:inspiration_in_DT}
    \Description{The figure shows the Double Diamond framework combined with IDEO’s three core design thinking activities. It presents four phases: Discover, Define, Develop, and Deliver, mapped onto three stages: Inspiration, Ideation, and Implementation. Discover is mapped onto Inspiration, while Define is shown with a gradient from orange to blue, indicating it spans both Inspiration and Ideation. Develop is mapped onto Ideation. Deliver is faded, signalling that Implementation is not considered in this work. Each diamond shape widens to represent divergent thinking and narrows to represent convergent thinking.}
\end{figure}

In the field of UX/UI design, solving complex problems and creating innovative designs using a human-centred approach is referred to as “design thinking”~\cite{HassoPlattnerInstituteofDesignatStanfordUniversity2024, IDEOU2020, interaction_design_foundation_what_2024, Rogers2023}. Researchers note that inspiration is essential to the design thinking process~\cite{cross2023design}. It allows designers to break away from existing ideas and explore unknown territories of ideas which could lead to innovation~\cite{Vasconcelos2016, Vinker2023}. Widely-used models of design thinking such as the Stanford D-School's 5 stage process~\cite{HassoPlattnerInstituteofDesignatStanfordUniversity2024}, and the Design Council’s double diamond process~\cite{DesignCouncil2024} do not incorporate the word "Inspiration" as a way to distinguish the stages in design process; therefore, in this work we use IDEO’s three core activities of design thinking – Inspiration, Ideation and Implementation~\cite{ideo2015field}– in discussing our work. According to IDEO’s model, in the \textit{inspiration} phase, designers become familiarised with the external or internal trigger(s)/sources that motivate or evoke new ideas. In the \textit{ideation} phase, designers actively generate and develop ideas for solutions. In the \textit{implementation} stage, designers make these ideas tangible, evaluating them with users, and validating which solutions work. Our design fiction focuses only on the initial two phases, Inspiration and Ideation. In Figure~\ref{fig:inspiration_in_DT}, we map IDEO's three core activities~\cite{ideo2015field} of design thinking on to the familiar Double Diamond framework~\cite{DesignCouncil2024} to illustrate how we situate our discussion.

Researchers believe that the more designers are exposed to inspiration stimuli, the more mental connections they make, leading to fluency in generating creative ideas~\cite{Andolina2015, Lallemand2024} and expanding their conceptual space~\cite{Boden2009, Chen2023}. According to Boden, conceptual space refers to abstract spaces in human mind, or styles of thinking, which prompt novel, valuable and surprising ideas~\cite{Boden2009, Chen2023}. Designers try to expand their conceptual space through exposure to different types of stimuli~\cite{Gonalves2016, Keller2006}. Conducting precedence studies and compiling Mood Boards~\cite{InteractionDesignFoundation2024, Koch2020, Lucero2012} by collating visual stimuli to identify trends, styles, and colours are some of the techniques designers use in the visual design stage in a UX/UI design project. However, exposure to inspirational stimuli may not always happen once a design project is started, therefore designers follow methods to store interesting stimuli, such as maintaining idea journals, creating photo albums and scrapbooks, downloading images from the internet and saving them in local folders, maintaining Pinterest boards, and creating online shared locations~\cite{inie2018designing, baigelenov2025visualization, rossellicreative2025}. When designers want inspiration for a project, they extract images and thoughts from these archives and create mood boards that may provide inspiration for ideation. Yet, managing these ideas become challenging due to accumulated material if the designers lack a proper system~\cite{inie2018designing, rossellicreative2025}.

In dual-process theory of ideation~\cite{gonccalves2021life}, researchers discuss that, while inspiration can act as a way to enhance creative thinking, it can become a negative influence unless it is used cautiously~\cite{Kulkarni2014}. Design fixation, which is known as “the blind adherence to a set of ideas or concepts limiting the output of conceptual design”~\cite[p. ~1]{Jansson1991}, is a negative experience faced by designers due to incorrect exposure to stimuli~\cite{Jansson1991,Crilly2019}. Prior literature discusses that inspiration depends on various factors such as: the modality of the stimuli, the time of exposure, the definition/fidelity of exposure, the distance between the design problem and the stimuli, the diversity of stimuli, the design process, the experience of the designer, the disciplinary background, the level of detail of the design problem, and instructions given the time available for the task~\cite{Vasconcelos2016, Leahy2018}. For example, designers tend to adhere to exposed ideas more strongly if they are concrete rather than abstract~\cite{Youmans2014a}. Researchers discuss different ways of mitigating the negative effects of stimulus exposure~\cite{Youmans2014b}. Changing the completeness of the stimuli~\cite{cheng2014new}, exposure to remote ideas~\cite{chan2011benefits}, time of exposure~\cite{tsenn2014effects}, and constant reminders~\cite{Youmans2014b} are discussed as remedies in design fixation literature. These findings emphasise that identifying the right amount of exposure to inspirational stimuli is essential.

\subsection{Generative AI in Design Workflows}

The recent boom in GenAI has yielded a number of AI-powered creativity support tools (AICSTs) developed by startups~\cite{Chui2023} as well as GenAI updates to existing creativity support tools (CSTs) from established vendors like Adobe~\footnote{\url{https://www.adobe.com/au/ai/overview.html}}, Autodesk~\footnote{\url{https://www.autodesk.com/au/solutions/autodesk-ai}}, Canva~\footnote{\url{https://www.canva.com/ai-image-generator}}, Miro~\footnote{\url{https://miro.com/ai}} and Figma~\footnote{\url{https://www.figma.com/ai}}. These AI tools are developed to support designers with various design activities, such as user research, ideation, prototyping, evaluation, and presentation. Apart from facilitating the core design tasks, there are GenAI solutions for project management and converting design into functional prototypes, and emerging approaches to using GenAI such as vibe coding and vibe design~\cite{uxTigersVibeCoding2025}. 

One common feature of all these tools is that they heavily depend on text prompting as their main input mode~\cite{khan2025beyond, Wadinambiarachchi2024}, which might frustrate users when interacting with these tools~\cite{Li2024}. The chat-based interface limits the ability for the users to move between iterations of a design except by scrolling through the chat log. Other problems include a lack of contextual awareness of AI tools, which leads to reiterations and a tendency of GenAI to produce stereotypical output~\cite{AgarwalHomogenize2025, Wadinambiarachchi2024}. Researchers discuss that, current AI tools follow rules and do not use true reasoning behind their action and GenAI is prone to hallucination~\cite{Kaate2025Hallucination}, thus the users must manually check outputs. Automated research insights may be shallow and cannot capture nuances, interpretiveness, and the depth a human can bring~\cite{thominet2024our, shneiderman2022human}.

Despite a lack of understanding of the long-term effects of GenAI on design work, professional designers of all disciplines incorporate GenAI into their workflows. For example, world-renowned Zaha Hadid Architects use Dall-E and Midjourney to develop design ideas~\cite{Barker2023}. Audi, a leading car manufacturer, has created a bespoke GenAI tool called FelGAN to inspire new rim designs~\cite{AudiMediaCenter2022}. Whilst many designers share their positive experiences of integrating GenAI tools into their workflows in blog posts and podcasts, prominent UX/UI organisations like Nielsen Norman Group argue that AI is not yet ready to take over design tasks~\cite{Sponheim2024, nngroup2025aidesign}.

In addition to the examples from industry given above, there has been a proliferation of HCI research work on designing novel AI tools, as well as understanding users’ adoption of AI tools into design work~\cite{choi2024creativeconnect, Rafner2023,Li2024, Uusitalo2024, Takaffoli2024}. These studies unveil various responses that indicate opportunities and potential caveats. Researchers have developed novel design tools including GANCollage~\cite{wan2023gancollage}, CreativeConnect~\cite{choi2024creativeconnect}, Expandora~\cite{choi2025expandora}, PromptInfuser~\cite{petridis2024promptinfuser}, Code Shaping~\cite{Yen2025CodeShaping}. Rafner et. al.~\cite{Rafner2023} studied AI’s potential to enhance creativity; on the other hand, Wadinambiarachchi et. al.~\cite{Wadinambiarachchi2024} observed higher design fixation and cause fixation displacement when AI-generated images were used as a source of inspiration stimuli in a rapid ideation task. Thus, we can assume that the way of interacting with the AI tools and the nature of the task may influence the quality of a designer’s experience. Additionally, HCI researchers also discuss inconveniences designers face when using prompting to communicate with AI tools which leads them difficulty attaining their desired outcomes~\cite{Li2024, Wadinambiarachchi2024}.

Even though the amount of academic and non-academic literature about AICSTs has exponentially grown, the existing research literature provides only a limited picture of how design professionals can use AI tools for managing inspiration. Our work addresses this gap by asking designers which activities in the inspiration and ideation stages of the design process should and should not be augmented by AI.

\subsection{Emerging Field of Agentic AI for Design Work}

Agentic AI is a new paradigm in AI, which refers to AI systems designed to operate autonomously~\cite{murugesan2025rise,HBRpurdy_2024,hughes2025ai}. Unlike prompt-driven GenAI tools, which respond only to users’ direct input, agentic AI systems can adapt to changing circumstances by analysing the environment~\cite{murugesan2025rise} and driving towards accomplishing a goal set by the user~\cite{hughes2025ai}. Agentic systems have evolved from executing an explicit command from the user, such as Amazon’s Alexa and Apple’s Siri, towards integrating multimodal inputs, providing context-sensitive suggestions and automating complex domain-specific workflows, and performing collaborations such as Microsoft Co-pilot~\footnote{\url{https://copilot.microsoft.com}}, Google’s Gemini~\footnote{\url{https://gemini.google.com/app}}, OpenAI’s Operator~\footnote{\url{https://openai.com/index/introducing-operator}} and Anthropic’s Claude~\footnote{\url{https://www.anthropic.com/claude}}. Currently, agents can utilise techniques such as Retrieval- Augmented Generation (RAG)~\cite{Lewis2020rag} (a method where AI retrieves external data to generate more accurate and contextually relevant responses), Model Context Protocol (MCP)~\cite{hou2025model} (a framework for ensuring agentic AI systems operate securely and contextually within specific environments), to deliver more accurate and contextually relevant responses to the user by using appropriate data retrieved from external sources~\cite{Jang2024au-RAG, ibm_2025AgenticRAG}. Furthermore, agentic systems can continuously learn from interactions and outcomes, and iteratively improve both the efficiency and quality of the work~\cite{murugesan2025rise, hughes2025ai}.

In design contexts, agentic AI systems are envisioned to help designers by breaking down complex tasks into manageable subtasks, dynamically adjusting strategies, and effectively managing a wide range of tools and external data sources to achieve desired goals more efficiently. Adobe, a leading design tool vendor, has unveiled its vision for enhancing creativity and productivity through agentic AI~\cite{greenfield_2025AdobeVision}. 

As this field is still in its infancy, empirical insights on how agentic AI will impact design workflows are scarce. Current AI tools are designed primarily for single users and rely heavily on textual prompts. Therefore, anticipating the entry of agentic AI into design tools, we aim to understand users' views, expectations and needs on integrating agentic AI into their workflows, allowing users to speculate how they want to interact with an A-AICSTs.

\section{Methods}

We conducted a design fiction study, one of the speculative design methods~\cite{Huang2023, Noortman2019, Sturdee2016} which is a subset of research through design approaches~\cite{Zimmerman2007}  used in HCI. Design fiction allows participants to envision future interactions~\cite{Blythe2014}. We created the design fiction “What could Idy do?”: a near-future scenario in which a group of designers use a design platform called "I-space" that incorporates an AI agent named "Idy" in searching for inspiration sources and identifying visual directions in a UX/UI design project (see section~\ref{sec:design fiction}). I-space includes features of both generative and agentic AI. We chose this method to provoke imagination on how a GenAI agent could support designers in the initial stages of the design process and how designers should interact with agentic AI in design workflows. We compiled our design fiction into a deck of “Flip-Flap story cards” to conduct a story card elicitation task (see section~\ref{sec:Flip-flap-cards}), which was followed by a worksheet completion activity. The study received human research ethics approval from both the University of Melbourne, Australia and University College London, UK.

\subsection{Study Design and Materials}

\subsubsection{Design Fiction: What could Idy do? }
\label{sec:design fiction}
We compiled our design fiction as follows. First, we imagined a real-world scenario in which a client approached a software development firm to design an app for tourists walking a hiking trail. Below, we describe the synopsis of the design fiction:

\begin{figure}[H]
    \centering
    \includegraphics[width=1\linewidth]{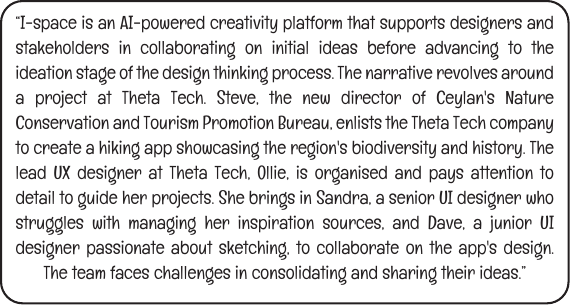}
    \caption{\textit{The synopsis of our design fiction "What could Idy do?"}}
    \label{fig:synopsis}
    \Description{The figure presents a text synopsis of the design fiction “What could Idy do?”. It describes an AI-powered creativity platform called I-space that helps designers move from inspiration to ideation. The narrative centres on a project at Theta Tech, where the team designs a hiking app and faces challenges in consolidating and sharing ideas.}
\end{figure}

Then, the first and third authors expanded the fictional narrative to a range of design-process scenarios that highlight the current practices and challenges designers face: from defining requirements, gathering inspirational sources, to sharing concepts, curating, structuring, and exploring stimuli, and finally synthesising and evaluating ideas for further design development. We represented the narrative as several distinct frames within a storyboard and wrote a question, "What could Idy do?" for each frame. To create possible responses to this question, the research team imagined ways an AI agent could intervene in each situation. Then, by utilising literary techniques such as metaphor, hyperbole, and irony, we elaborated on potential solutions on how Idy, the AI agent in I-space, could support designers in the design process.

\begin{figure*}[htbp]
    \centering
    \includegraphics[width=0.9\linewidth]{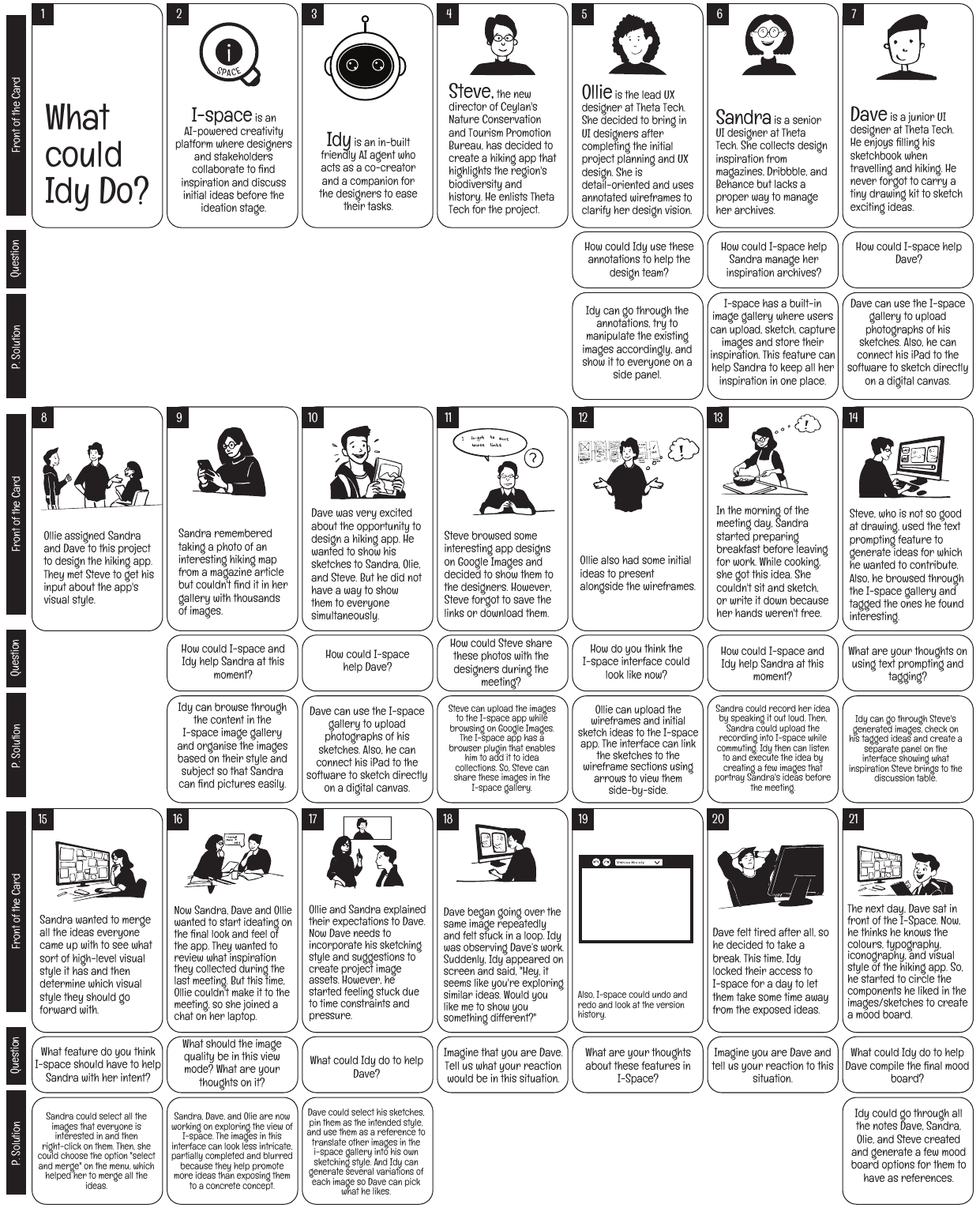}
    \caption{  \textit{``What Could Idy Do?”} design fiction, Flip-Flap story cards (some with questions and suggested answers) presented to the participants}
\label{fig:story_cards_front}
\Description{The figure shows the complete set of Flip-Flap story cards used in the design fiction “What Could Idy Do?”. Each card combines illustrations with text, including character descriptions, scenario prompts, guiding questions, and suggested answers. The story introduces I-space, an AI-powered platform, Idy an AI agent and the characters Steve, Ollie, Sandra, and Dave, who collaborate on a design project using the tool. Participants used these cards to explore how Idy and I-space could support different stages of their design work, such as gathering inspiration, organizing ideas, and resolving challenges.}
\end{figure*}

We represented the story frames as a set of prototype interactive cards and conducted pilots with three HCI Graduate Researchers. These pilots helped us decide which questions to retain and refine, which solution prompts to include for critique, and how to condense the narrative to fit a one-hour session. After finalising the Flip-Flap story cards, we invited participating designers to reimagine, critique, and contribute their perspectives to the story flow (see Figure~\ref {fig:story_cards_front}).

\subsubsection{The Flip-Flap Story Cards and the Worksheet}
\label{sec:Flip-flap-cards}
\begin{figure}[H]
    \centering
    \includegraphics[width=1\linewidth]{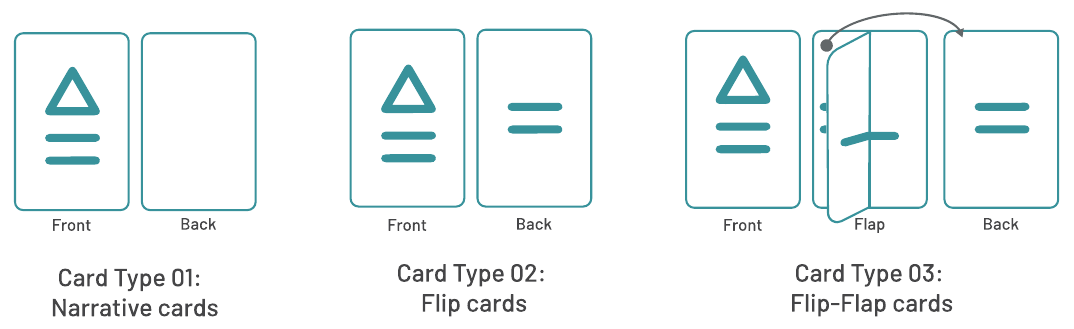}
    \caption{ Types of cards used in the deck of Flip-Flap story cards. Card type 01: Narrative cards, Card type 02: Flip cards, Card type 03: Flip-Flap cards }
\label{fig:flip_flap_cards}
\Description{The figure illustrates three types of cards in the Flip-Flap story card deck. Card Type 01, Narrative cards, provide context to participants. Card Type 02, Flip cards, show a story frame with an illustration on the front and a related question on the back. Card Type 03, Flip-Flap cards, are similar to Flip cards but include an additional flap between front and back, used to reveal a hidden potential solution.}
\end{figure}

The Flip-Flap story cards were inspired by a magician’s three-sided trick cards as well as card-based research approaches conducted in HCI~\cite {Halskov2006, Roy2019}. The deck of cards had three types of story cards (see Figure~\ref{fig:flip_flap_cards}), each serving three different purposes.

\begin{itemize}
    \item \texttt{Narrative cards} provided context to the participants.
    \item \texttt{Flip cards} contained a story frame with an illustration on the front of the card, and a question at the back.
    \item \texttt{Flip-Flap cards} were similar to Flip cards, which contained a story frame on the front and a question at the back, but included an additional flap that could be opened to reveal a potential solution.
\end{itemize}

\begin{figure}[H]
    \centering
    \includegraphics[width=1\linewidth]{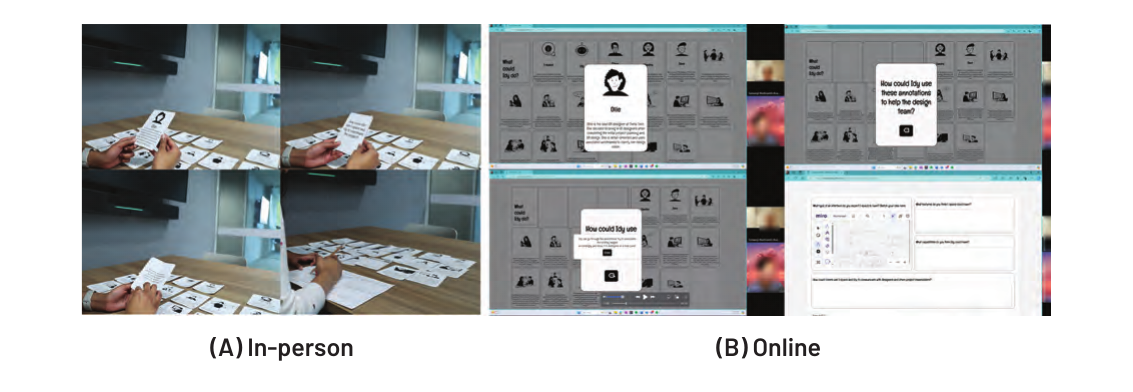}
    \caption{  Designers participating in one-on-one sessions.    (A) In-person sessions, (B) Online sessions}
    \label{fig:oneONone}
    \Description{This figure illustrates a four-step activity sequence, presented both in-person and online. In the in-person sessions, participants first read the description on the front of a physical card, then flip the card to read the question on the back and provide a response. Next, they open the flap to reveal the hidden solution. After the participants finished the elicitation activity by going through all the story cards, they completed the worksheet on paper. In the online sessions, participants follow the same sequence using digital equivalents: reading the description on a virtual card, opening the back of the card to view the question and type a response, clicking to reveal the hidden solution, and filling in a digital worksheet within the interface.}
\end{figure}

The flip-flap mechanism (see Figure \ref{fig:flip_flap_cards} - Card type 03) was designed to allow participants to first develop solutions based on their own experience, and then critique solutions suggested by the research team. We created both physical and digital cards so that we could conduct the study in-person (see Figure~\ref{fig:oneONone}-(A)) and online~\ref{fig:oneONone}-(B)). Digital cards presented the solution in a pop-up that appears after clicking a button (Link to~\href{https://samangiwa.github.io/what_could_idy_do/cards.html}{digital story cards}). Finally, we provided participants with a worksheet (see appendix~\ref{Appendix_A}) to complete after the elicitation task. This allowed them to summarise thoughts gathered while going through the story, and then imagine beyond the storyline.

\begin{figure*}[htbp]
\centering
\includegraphics[width=1\linewidth]{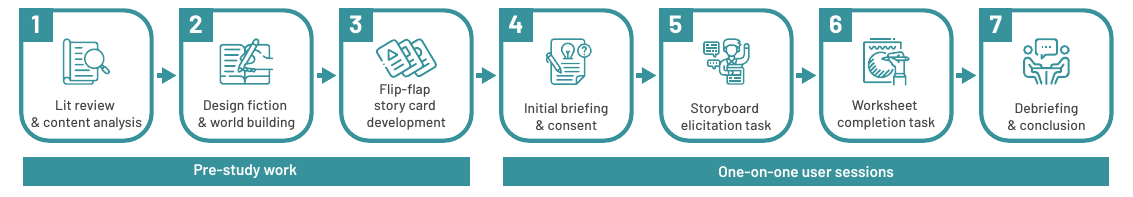}
\caption{  Overview of our study method, structured in seven steps. (1 -3) Pre-study work including (1) Lit. review and content analysis, (2) Design fiction and world-building, (3) Flip-Flap story card development, (4 -7) One-on-one user sessions including (4) Initial briefing and consent, (5) Story card elicitation task, (6) Worksheet completion task, (7) Debriefing and conclusion.}
\label{fig:study_flow}
\Description{The figure shows a seven step study process organized into two phases. Pre-study work includes: (1) literature review and content analysis, (2) design fiction and world-building, and (3) Flip-Flap story card development. One-on-one user sessions include: (4) initial briefing and consent, (5) a storyboard elicitation task, (6) a worksheet completion task, and (7) debriefing and conclusion. The steps are displayed in sequence with arrows indicating progression.}
\end{figure*}

\begin{table*}[htbp]
    \centering
    \caption{  Participant information}
    \small{
    \begin{tabular}{p{0.5cm} p{0.75cm} p{2.5cm} p{0.75cm} p{1cm} 
                p{0.75cm} p{1.25cm} p{1cm} p{0.85cm} p{1.5cm} 
                p{0.5cm} p{0.5cm} p{0.5cm}}
\toprule
\textbf{PID} & \textbf{Pseud.} & \textbf{Profession} & \textbf{Prof. Exp (yrs)} & \textbf{Emp. Status} & \textbf{Country of Emp.} & \textbf{Gender} & \textbf{Age} & \textbf{Highest Edu.} & \textbf{Major} & \textbf{AI Exp (yrs)} & \textbf{Attitude towards AI}  \\ 
\midrule
P01 & Aaron & Design Researcher & 4 & Full time & UK & Male & 27 & PhD & HCI & 1< & + \\

P02 & Ben & Postdoc-researcher & 1.5 & Full time & UK & Male & 30 & PhD & HCI & 2< & + \\

P03 & Clara & UI/UX designer & 9 & Full time & UK & Female & 35 & Masters & Interaction Design & 1< & +\\

P04 & Dylan & Design Researcher & 4.5 & Full time & UK & Male & 29 & Masters & Liberal Arts & 1< &  - \\

P05 & Elton & HCI Researcher & 12 & Full time & UK & Male & 37 & PhD & Design & 1< &  + \\

P06 & Frank & Design Researcher & 20 & Full time & UK & Male & 33 & Masters & Design Eng. & 2< & + \\

P07 & Gaia & Design Researcher/ PhD Cand. & 4 & Full time & UK & Female & 37 & Masters & Design Res. & <1 & + \\

P08 & Hailey & Project manager (Communications) & 12 & Full time & UK & Female & 67 & Masters & Linguistics & <1 & 0 \\

P09 & Isabel & Product Designer & 5 & Full time & UK & Female & 33 & Masters & UX Design & <1 & + \\

P10 & Julia & Design Researcher & 3 & Full time & UK & Female & 25 & Masters & Design & 2< & ++ \\ 
\bottomrule
    \end{tabular}
    }
    \label{tab:participants}
    \Description{The table provides demographic and professional information for ten study participants, identified by pseudonyms. Participants include design researchers, HCI researchers, product and UI/UX designers, and a project manager. All were employed full-time in the UK, with professional experience ranging from 1.5 to 20 years. Ages range from 25 to 67, and education levels include Master’s and PhD degrees across fields such as HCI, interaction design, design engineering, linguistics, and liberal arts. Most had limited prior experience with AI (less than two years), and their attitudes toward AI ranged from positive to neutral or negative.}
\end{table*}

\subsection{Participants}

We recruited 10 professional designers (5 men and 5 women, see Table~\ref{tab:participants})~\footnote{gender was self-described by participants} through mailing lists, social media/online design communities, and word of mouth. Participants scheduled their sessions through a digital sign-up form. Before the session, we emailed them an instruction sheet, consent form, and demographics questionnaire. Participants self-reported their prior design experience in years/months. All participants were 18 years or older and we assigned pseudonyms to participants to ensure their anonymity. The mean age of the participants was 35.3 years (25–67, SD = 11.85).

\subsection{Procedure}
The sessions were conducted individually, either in-person in the participants’ studios, on university premises, or online (via Zoom or Microsoft Teams), depending on participants’ preference and availability. 6 participants joined in-person sessions while 4 joined online. We began each session by explaining to participants how the study was being conducted and how the cards worked (see Figure \ref{fig:study_flow}- stage 4). During the story card elicitation task, participants were asked to think aloud~\cite{McDonald2012, Obrien2023} (see Figure \ref{fig:study_flow}- stage 5). Participants read the story on the front of each card and responded to the question on the back of the card. In the case of Flip-Flap cards, we told participants that there was a potential solution in the flap of the card and asked them to critique it after providing their own solution. Once all the 21 story card elicitation tasks were done, participants were asked to complete a worksheet (shown in appendix~\ref{Appendix_A}). The entire procedure lasted a mean of 53.70 minutes (44-60 mins, SD = 5.89). Finally, we reimbursed participants for their time with a \$25 (AUD) gift card.

\subsection{Data Analysis}
Our data included \textit{audio recordings} from the story card elicitation task and the \textit{worksheets completed by the participants}. Audio recordings were automatically transcribed by Microsoft Word, and the transcripts were verified and corrected by the first author. We followed Braun and Clarke’s 6-phase reflexive thematic analysis approach~\cite{Braun2022, Terry2021}. The first author coded the transcripts using an inductive data-driven approach. These codes were then grouped to develop \textbf{12 tentative themes}, which captured insights into tasks supported by AI, ways of expressing intent to AI systems, and how work and authority could be allocated in design workflows. Through iterative team discussions, we refined these themes by resolving overlaps, removing redundancies, and interpreting insights beyond the surface-level meaning. This process resulted in \textbf{8 themes}, which we further clustered into \textbf{3 overarching themes} that synthesised and extended the initial themes.

\section{Findings}

Through our analysis, we generated three overarching themes: 
\begin{itemize}
    \item Theme 1: Potential roles for AI agents in finding inspiration for ideation (responding to RQ1).
    \item Theme 2: Distribution of labour and authority among human designers and AI agents (responding to RQ2). 
    \item Theme 3: the need for better human-AI communication that extends beyond prompts (responding to RQ3).
\end{itemize}

In this section, we elaborate on the themes and sub-themes we identified. In the following sections, we label quotes from participants using the notation: “SC-x: A”, “SC-x: C”, where “SC-x” indicates the story card number, “A” indicates the participants’ answer to the question at the back of the card, “C” indicates the participants’ critique of the potential solutions we provided on the flap. We label worksheet responses by the participants as “W”.

\subsection{Potential Roles for AI Agents in Finding Inspiration  for Ideation}

\label{sec:Theme1}

Discussions revealed the different roles participants believed AI agents could perform in agentic AI systems. Participants saw the AI agent as a \textbf{\textit{1) work coordinator, 2) resource steward, 3) reframer, 4) guardian, and 5) creative catalyst.}}

\subsubsection{Agent as Work Coordinator.}
\label{sec:Theme1-1}
Like the designers portrayed in the design fiction scenario, participants acknowledged that they often experience difficulties in articulating ideas, organising thoughts and resources, and communicating with stakeholders. As a solution, participants suggested that an AI agent could act as a \textit{"coordinator"} or \textit{"moderator”} (SC 5: A- Aaron) who can \textit{"create meeting summaries and feedback"} (W-Elton) and \textit{"create task lists"} (W-Dylan) to enable collaborators to \textit{"... not lose focus on what that goal is"} (SC 15: C- Dylan).

Participants discussed how AI agents could provide individual and team-level action points to guide all the stakeholders in a project. For example, Frank said AI could: \textit{“maintain a list of all the various projects that an individual is working on, so you can have a system that looks at a project team in different ways, could be at the individual level, could be at the team level”} (SC 6: A- Frank).

Julia suggested that \textit{"[AI agents could] monitor and provide on-the-spot support... yet with a proper strategy”} and enable stakeholders to use AI to \textit{“auto-generate meeting minutes to update senior leaders in the firm”} might be a good way of utilising AI in a collaborative setting (W-Julia).

Similarly, participants suggested that AI agents could support designers by planning, organising tasks, creating timelines, and providing technical assistance, in which they could break through operational blockers encountered in design projects such as overwhelm, time constraints and external pressure:

\begin{quote}
 \textit{``... if he’s [the designer is] feeling stuck, if he’s struggling with time constraints and pressure, you could offload that to Idy who could act more as a planner and create a timeline of what he needs to do and how, so that would be like more logistical planning support. But also, it would have that technical support for feeding images and supporting the technical aspects.''} (SC 17: A- Aaron)
\end{quote}

These responses suggest that \textit{\textbf{AI agents can support designers by coordinating them with plans and guidelines to keep them on track towards the goal.}}

\subsubsection{Agent as Resource Steward.}
\label{sec:Theme1-2}
In response to scenarios where story characters experienced information overload, inconsistent methods of managing and capturing ideas and resources, and difficulties retrieving them, participants indicated that agentic AI could create space for high-level thought by automating mundane activities. These activities included renaming, organising, and categorising design assets and making it easier to find and manage files. For instance, Clara said that:
\begin{quote}
\textit{``I think it would be easy for a person to scan and store their information in one place and get AI help to rename it properly. So, it’s not like wasting time to rename and sort the information''} (SC 7: A- Clara).
\end{quote}

Participants suggested AI-powered tools could tag and categorise information and assets based on visual similarities, help designers to surface this information, and use existing asset metadata to make classifications:
\textit{"[AI agents should be] able to catalogue visual inputs by image recognition"} (W-Hailey)
\textit{“If Sandra took the photo and then uploaded it with metadata, the tool could help in the search process. Maybe the AI is able to infer some kind of object, activity, place, features or some of these images. So, if they search for a map or let’s say hiking, for example, presumably the number of images you will look through would reduce by a lot”} (SC 9: A- Dylan).

Extending this notion further, responding to story card 6, on which Sandra has difficulties managing her archive of ideas, Aaron suggested that:

\begin{quote}
    \textit{“[The AI agent] would support and provide the scaffolding to do this management exercise, rather than just doing it automatically, and by doing that, it allows the designer to find inspiration by linking stuff together she wouldn’t have thought of putting together. So, maybe there’s even bits where it’s purposefully put in the wrong bits, and maybe that crossover, like pollination, can be a good thing, but with the end goal of, supporting this management task”} (SC 6: A- Aaron)
\end{quote}

This indicates opportunities for collaborative exercises, where AI agents sometimes purposefully give wrong information to keep the designer cautious and mindful of their tasks. This could be a strategic way to reduce the over-reliance on AI systems. These responses imply that \textit{\textbf{AI agents can assist designers by playing a semi-autonomous role, or a cooperative role with the users.}}

\subsubsection{Agent as Guardian.} 
\label{sec:Theme1-3}
Participants suggested that AI could \textit{"Not just [act] as an assistant but also [provide] scaffolding [for design tasks] in personalised ways."} (W-Ben). AI agents could maintain \textit{“contextual awareness”}, meaning the AI agent could continuously monitor and track the ideas explored when accessing sources of inspiration and prevent designers losing track and missing information. Participants imagined this AI agent could auto-save content and browsing histories, while making it easy to surface information later via queries.

\begin{quote}
    \textit{``Idy [AI agent] has awareness of what he [the designer] was browsing for. It doesn’t mean that Idy would have to monitor everything you do, but maybe whenever you are working on the project and making searches related to the project, you can set it to do where everything I do in this window is related to that design project,... So, you could potentially then just ask it [AI agent]: ‘There was this thing that I saw a few days ago about XYZ, and then it can surface...''}(SC-11: A- Ben)
\end{quote}

Participants suggested taking care when making AI systems contextually aware, e.g.: \textit{``Ethically it could be a problem, but maybe you can give permission to do that''} (SC-11: A- Aaron). Allowing users to select what they want to be tracked is essential. Frank suggested:
\begin{quote}
\textit{`` ..maybe there could be something like \textbf{a mode} where, OK, now I’m going to look at things that are related to this project, and so you [permit agent to] just capture everything that is happening right now''} (SC-11: A- Frank)
\end{quote}

Participants suggested that guardian agents could use checklists and checkpoints generated by coordinator agents to ensure designers are on track, verifying that their activity is within guardrails. \textit{``[AI agent could] use it [checklists] as a way to help me stay on task, but I would not make it do the task components.''} (SC 15: C- Dylan)

Participants reported finding long design processes daunting, leading them to forget options or ideas as they proceeded. They indicated that version control in current GenAI systems is limited, and highlighted the importance of utilising better strategies, in-context learning and greater computational powers to perform better version control of ideas.
\begin{quote}
\textit{`` Version history is always good, ... Current [AI] doesn't do a very good job at version history and there's no way of tracing back to time points, ... I think there is no way of properly navigating through these branches in current popular [AI] tools, but if [that existed], I would use it''}(SC 19: A- Elton)
\end{quote}

Participants suggested that AI tools could adapt to a designer’s style and preferences over time, becoming more aligned with the designer’s creative goals. This may allow AI to offer suggestions that fit the designer’s evolving style, expanding the conceptual space in personalised directions. They suggested that \textit{``...[If AI] is able to record and eventually remember the things that you’re doing right now. That might be helpful for a later generation of new ideas based on whatever you were browsing...'}' (SC-11: A- Frank). That is, AI agents could guide designers by reminding them what they were previously browsing and providing alternative ideas and suggestions based on their past search.

However, Ben suggested that AI agents may \textit{``show what could be good to do, but this would probably only be in terms of suggestions, or in a way that you kind of peripherally get this information.''} (SC 20: A- Ben). The designer should therefore have the opportunity to take action rather having tasks forced on them by AI agents.

These responses indicate that designers envision that \textbf{\textit{AI agents could support designers by maintaining contextual awareness, learning from designers’ style and preference, and guiding designers to stay on track.}}

\subsubsection{Agent as Reframer.}
\label{sec:Theme1-4}
Participants imagined that reframer agents could nudge designers to recall what they might have missed, and suggest new trajectories, to help designers reframe their ideas. For example:
\begin{quote}
\textit{``The machine can generate multiple options, but then the humans can use that as a way to tease out what exactly they liked about previous versions, what are some things they really want to keep in the new design?...''} (SC 15: A- Julia)
\end{quote}

Frank suggested that an AI agent could provide a point of view that did not otherwise exist within the designer team, providing additional ideas and counter-examples to prompt designers to explore alternative solutions:
\begin{quote}
\textit{``Rather than just acting as a tool and a completely neutral party, it could come in with either counter examples, which I think could be really helpful in the very early stages,... where you want to have divergent ideas at the start, so offering alternative viewpoints on how you are going about the process...''} (SC 5: C- Frank)
\end{quote}

Ben suggested that agents could recommend ideas or sources of inspiration in the designer’s peripheral vision:
\textit{``This would probably only be in terms of suggestions, or in a way that you kind of peripherally get this information.''} (SC 20: A- Ben)

These responses denote that participants envisioned \textbf{\textit{AI agents assisting them in recalling missed ideas, exposing to counter examples during their ideation process and support designers reframing their concepts, guiding them to explore new directions without being intrusive.}}

\subsubsection{Agent as Creative Catalyst.} 
\label{sec:Theme1-5}
Participants suggested agents could be catalysts for novel thinking. Agents can \textit{"generate some ideas and those can be jumping off points for Dave [designer] to iterate upon.”} (SC-17: A- Frank), Participants suggested this could be implemented through view modes:

\begin{quote}
\textit{``Let’s say I only want to see [ideas from] Dave [junior designer] and Sandra [senior designer] side-by-side now, or I want to see Dave and Steve [client], or I want to see Sandra and Steve, [...] like pair-wise or like sets of three. ''} (SC 15: A- Dylan)
\end{quote}

Gaia suggested that AI could identify commonalities across the design team’s work so that individuals can easily recognise what is important:
\begin{quote}
\textit{``Idy could give some insight of the commonalities between the ideas. So, Sandra [designer] could see what the thread is, what is the commonality between the people’s collected content, that means things that should be kept.''} (SC 15: A- Gaia)
\end{quote}

Participants preferred low fidelity and abstract ideas in early stages of design: \textit{“lower fidelity allows people to feel more comfortable to critique}” (SC-16: C -Julia), whilst at later stages, visuals should be high fidelity while designers choose a final direction:
\begin{quote}
\textit{``Because the decision should be made based upon the high-fidelity versions of things, assuming that it is about the visual aesthetic at this point''.} (SC-16: C -Frank)
\end{quote}

Dylan suggested AI agent could narrow down the options from a broad range to a smaller set to support decision making: \textit{``...low fidelity images can promote more ideas, but as I’m trying to move towards something, I think it’s worth pushing people to say hey, looks like we actually have these three options.''} (SC-16: C -Dylan). Extending this idea further, Frank suggested that AI seamlessly switch between modes, proposing “AI-powered lenses” that allow designers to switch between different fidelities, styles, or perspectives as required: 

\begin{quote}
\textit{``If you're deciding like the final things and it's not the initialisation stage of sketchy stuff...it should be a fairly high high-fidelity representations of things. But with the ability potentially switching to low fidelity if you want to go back to sketching and suggesting new things...''}(SC 16: C- Frank)
\end{quote}
These responses reveal that \textbf{\textit{AI agents could enable designers to switch between different fidelities as they move between different phases of ideation and in the design thinking process.}}

 Overall, these responses indicate \textit{multiple potential roles for AI agents in finding inspiration in ideation}. The the degree of collaboration of the supportive role is task-dependent, implying that AI agents should be adaptable and flexible. Participants envision AI agents’ involvement to vary between active and passive roles, based on cognitive complexity whether it is mundane, analytical or creative task, and with varying degrees of creative agency and responsibility. However, they felt the ultimate decision making should remain in designers’ hands. 

\subsection{Distribution of Authority Among Human Designers and AI Agents}
\label{sec:Theme2}
Participants were wary of ceding control to AI and of giving clients access to the same AI tools that they used. Most preferred AI to tackle routine, low-risk chores with limited creative independence.

\subsubsection{Designers Should Be the Main Actors in Design Workflows.}
\label{sec:Theme2-1}

Participants believed humans should retain the authority to make creative decisions. Responding to story card 13, in which Idy automates a process when Sandra was unable to write her ideas because her hands weren’t free, Dylan felt: 

\begin{quote}
    \textit{“I’m not comfortable with the last point of Idy can listen to it and execute, I think that especially at a stage like this [development stage], I would want to be in control”} (SC 13: C- Dylan)
\end{quote}

Responding to story card 15, Isabel asked, \textit{“Can human beings really trust AI to do this [merging ideas] creatively and powerfully?”} She was concerned that AI may not be able to capture the visual essence and taste implicit in discussion. Julia noted, \textit{“I would be wary of just relying on the tech to merge [ideas] because again, it's a very stochastic model”} (SC 15: C- Julia). Similarly, Dylan expressed concerns that AI would be unable to capture nuanced human experiences, highlighting tensions about trust in human-AI interactions:

\begin{quote}
\textit{“I would not trust the AI to capture the diversity of those different ideas and what each of those ideas meant, or how each of those ideas emerged. The AI would not be able to capture all of the soft conversations or discussions”} (SC 15: C- Dylan)
\end{quote}

Responding to another story card, Dylan felt AI should help designers in times of need, but not decide:

\begin{quote}
\textit{“It [AI agent] could help you to stay on task and empower you as the designer or the UI person or the UX person, at each stage and be a kind of get-out-of-jail when you’re really stuck. Help you come up with some stuff, and then you have control and choose.”} (SC 21: C- Dylan)
\end{quote}

Similarly, Ben said the AI agent should \textit{"Not [give] a definitive answer, always leaving the last decision to the human."} (W-Ben)

Participants further noted that an archival tool should be owned by the designers themselves: \textit{“An inspiration archival tool is useful as long as it’s user-controlled, user-generated, and user-owned” }(SC 6: C- Dylan). 

Responding to story card 20, where Idy locks the system down for the day, participants felt that the level of autonomy given to AI should depend on the stage of the design process and the criticality of the task. For example, during the early exploratory phases, designers are more open to suggestions, whereas in later stages, they prefer more control. Julia said:

\begin{quote}
\textit{``If it’s a low-stake context where I’m just really exploring playing with different ideas and then yeah, like Idy to surface whatever it wants because I’m not under pressure, I’m open to inspirations... But if I have a specific goal I need to achieve and I have a plan on how I’m gonna achieve that like I have an idea and to execute it and then it [AI agent] shows me, here’s some fun ideas for you to explore. I’m like, NO, I don’t want this, it’s high-pressure, high-stakes situation. So really depends on the context.''} (SC 20: A- Julia)
\end{quote}

These findings suggest that \textbf{\textit{even though participants envision coexistence with AI agents, they believe AI does not have the requisite empathy, cultural sensitivity, and sensibility to make appropriate creative decisions. Thus AI should not be fully autonomous, and the autonomy given to agents should be context and task dependent.}}

\subsubsection{Use of AI by the Client and Other Stakeholders.}
\label{sec:Theme2-2}

Participants acknowledged that the client should be invited to contribute by providing feedback on creative decisions. Some participants said they would not want clients to have access to an AI-powered creativity platform, and that clients should only give suggestions and feedback.

\begin{quote}
\textit{“I wouldn’t want clients to use I-space. [It would be better to have a] separate final design space to share with client, so they can comment but not change the design or Idy could help to give them suggestions which is align to the final design themes or aesthetics.”} (W- Gaia)
\end{quote}

Other participants highlighted that clients' access to an AI agent should be provided with care, \textit{“making sure there’s a clear section in the platform for the client and providing them with the right level of authority (balanced)”}. (SC 21 – C Aaron)

Similarly, Isabel felt that: \textit{“They [the client] should be able to see the latest design files (not able to edit it) and leave comments around the design. The comments should be left at the same space and easier for everyone to review the comments”} (W- Isabel) 

Most participants endorsed having a separate view tailored for clients: \textit{“you have a panel which shows you and can be the base for a conversation with the client.” }(SC 14: C- Dylan)

However, some participants preferred having a space where the client could give contextual information and needs to the AI, which could then act as a mediator for sharing those ideas as suggestions:

\begin{quote}
\textit{“I guess there could be his own part or profile or part of the platform for the client who’s able to then share their own ideas, right? Maybe he could immediately talk to Idy and describe what the app designs looked like”} (SC 11: A- Aaron)
\end{quote}

These responses suggest that designers want to update clients on progress, but do not want clients to intervene in the creative process. Thus in most cases, \textbf{\textit{client-facing AI agents should act only in the space where the designers’ work is presented to clients}}.

\subsection{Effective Ways for Designers to Express Their Intent to AI Agents}
\label{sec:Theme3}

Although designers expressed a desire to interact with AI agents, they said that one of the bottlenecks for making these interactions meaningful is the difficulty of prompting. Current literature on AICSTs also discusses this as a concern and a limitation~\cite{Wadinambiarachchi2024, Li2024, Uusitalo2024, Takaffoli2024,Verheijden2023}. Our design fiction portrayed a variety of input techniques including annotations, sketches, text prompts, and voice commands, encouraging participants to imagine ways of interacting with AI agents.

Participants believed most designers are visual thinkers, leading to friction in the interface when designers must express their original imagination in words. 

Participants discussed a number of alternative modalities. Some were extensions of existing prompting capabilities. Aaron and Hailey suggested that AI could guide the user to structure their prompt, providing \textit{“scaffolding”} or \textit{“a taxonomy”} or asking questions of the user. 

Frank felt that rather than providing a blank text box to write prompts, AI should provide guided instructions to help the designer ideate:

\textit{“...getting started with writing is not always the easiest... But AI could start out with auto-generating things for you and giving you like pre-filled text for OK, here’s an idea that flows out of the text box...”} (SC 14: A – Frank)

Participants also nominated \textit{annotations} - adding explanatory notes or comments to visual content - as an approach to providing unambiguous input to AI. The AI agent can read the annotations and learn users’ intentions regarding specific parts of the content. For example, Isabel suggested: \textit{“I can just circle all the post-its now and ask them to do the summaries.”} (SC 19: A- Isabel)

Annotations could provide contextual information, allowing AI to find and pull together similar work or content:

\textit{“Idy could scan the annotations, understand what kind of user workflow this would be, and whether there were similar ones that were created before, so Idy could sort of pull together similar designs from the past and surface that to the team to serve as inspiration.”} (SC 5: C- Julia)

Elton further noted: \textit{“With text, I think there’s still some disconnection, like there’s a bit of a gap between you being able to imagine something and it being able to produce what exactly you’re imagining. So, I feel there’s something in between that you could do on top of text.”} (SC 14: A- Elton)

Ben suggested using \textit{sketching} to convey to AI the look and feel of an idea: AI could use these sketches as a starting point to generate more versions as inspiration:

\textit{“Dave [the designer] could provide a few sample images to give the AI an idea of how it should look and feel like, and then continue with more basic sketches that the AI then builds upon, and create a more refined version. AI then uses whatever Dave gave it as training.”} (SC 17: A- Ben)

Participants felt that \textit{tagging} could help AI agents to understand their ideas and intentions:

\textit{“I think tagging is useful. It helps, I think, especially in a collaborative setting, just like it can help other people very quickly get a sense of what you like.” }(SC 14: A- Dylan)

Hailey suggested using standard rather than custom tags to inform AI about common intentions:

\textit{“Because they’re working as a team and these people belong to a design company, I think it would be very interesting to have a common of standard tags which people then can use in a standard way, rather than creating hundreds of different tags, which then become too many to handle.”} (SC 14: A- Hailey)

Participants also suggested using \textit{voice} to interact with AI. Dylan shared his practice of using voice notes to archive his ideas:

\textit{“... I have a WhatsApp chat with myself, where I often talk to myself and I put stuff there and, then, refer to it later.”} Inspired by this practice, he suggested,\textit{ “You can talk to Idy through voice, talk to AI and then the AI is making notes for you as you’re talking or trying to interpret what the intent is.”} (SC 13: A- Dylan)

Aaron suggested that AI could transcribe the designer's voice inputs to annotate the content the user is pointing at:

\textit{“As you are speaking, it [AI agent] begins to annotate the aspects of the wireframe she maybe points at or she’s talking about, and the system is able to know that … it begins to generate, you know, whatever she’s speaking, but as an annotation.”} (SC 5: A- Aaron)

Responding to story card 5, Clara suggested using voice to command AI, \textit{“What if this AI platform can do a task if we give a command like okay, can you sort this out properly”}.

\begin{figure}[htbp]
    \centering
    \includegraphics[width=1\linewidth]{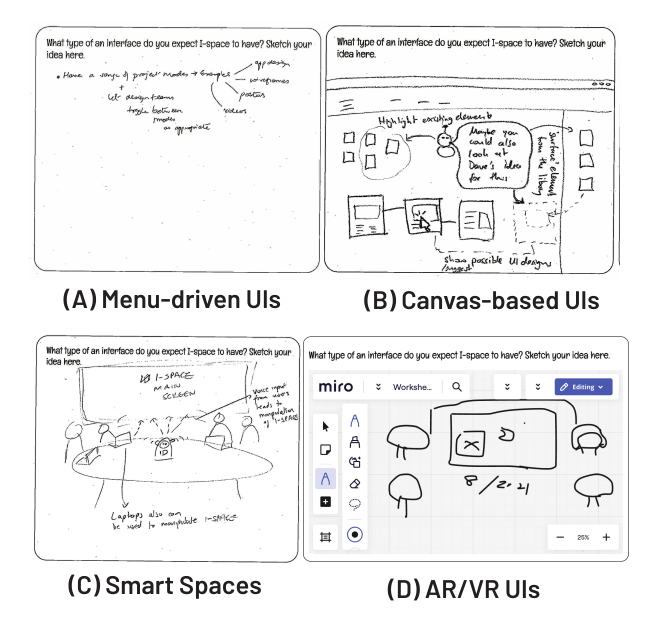}
    \caption{  Interface types suggested by participants: (A) Menu-driven UIs, (B) Canvas-based UIs, (C) Smart Spaces (Inspired by spatial and audio interfaces) and (D) AR/VR-based UIs.}
    \label{fig:interfaceTypes}
    \Description{The figure presents four types of interface concepts sketched by participants. (A) Menu-driven UIs show structured lists and options for navigating content. (B) Canvas-based UIs depict a free-form space where elements can be arranged visually. (C) Smart Spaces illustrate a physical meeting room with people interacting around a shared table, inspired by spatial and audio interfaces. (D) AR/VR UIs show a digital whiteboard with virtual objects and participants represented in an immersive environment.}
\end{figure}

Participants also discussed \textit{existing user interfaces} that they thought could be used to interact with AI (see Fig.~\ref{fig:interfaceTypes}). They sketched these ideas during worksheet completion. We categorised them as:

\begin{itemize}
    \item Interfaces inspired by tools like Adobe Photoshop that use menus and direct manipulation techniques.
    \item Interfaces inspired by existing infinite canvas-based design tools.
    \item Interfaces inspired by audio-controlled smart spaces like Google Home or Alexa.
    \item Interfaces inspired by immersive technologies such as AR and VR.
\end{itemize}

Collectively, these findings emphasise that participants want multimodality when sharing their intent with agentic AI systems. They suggested \textbf{\textit{annotations and sketching, non-verbal cues or verbal commands, to provide the context for content, and to provide directions for AI to act upon}}.

\section{Discussion}

This study aimed to understand how designers envisage AI agents supporting their workflow, particularly in gathering inspiration, expanding their conceptual space, and ideation. \textbf{Our findings highlight that, to integrate agentic AI into design workflows responsibly, it is crucial to identify different roles in design workflows, strike the right balance in distributing authority among AI agents and human designers, and facilitate seamless communication beyond prompts.}

In this section, we reflect on learnings from our study and highlight the importance of adapting user-centric approaches when designing agentic AI systems. We start by relating findings to prior literature, and drawing on the key themes from our findings to propose design directions, discussed in section~\ref{Disc:pt1}. Then, we synthesise our findings into a framework for designing interactions and boundaries between human designers and AI agents, in section~\ref{Disc:pt2}. Finally, we reflect the on the ``Flip-flap story cards method" in section~\ref{Disc:pt3}, the limitations and future directions of our study in section~\ref{Disc:pt4}. 

\subsection{Design Directions for Agentic AI-powered Creativity Support Tools}
\label{Disc:pt1}

AI is currently being incorporated into existing CSTs, and new AI-based tools are emerging that cater to most tasks in each stage of the design process~\cite{Takaffoli2024, Palani2024}. However, designers and commentators note that current AI tools are not ready to take over the design workflow as vendors have promised~\cite{Sponheim2024, nngroup2025aidesign}. Nevertheless, these tools keep evolving and may one day become ready to perform the intended tasks. Thus, an important question is “How could we integrate AI into design workflows responsibly to empower designers while maintaining the creative control in the human designer's hands?”.

\subsubsection{Designing Agentic AI Systems Where Agents Perform Different Roles in Design Workflows.}

Our participants indicated different roles AI agents could play in design workflows: \textit{work coordinator, resource steward, reframer, guardian, and creative catalyst} (see section~\ref{sec:Theme1}), and highlighted the importance of having a critical stance when incorporating AI agents into design workflows. 

Participants wanted context-aware AI agents that could support them in managing their inspiration archives and guide their collaborative design processes. For example, AI might mix different ideas in ways that human designers might not typically think of, by utilising creativity techniques such as remote associations~\cite{ChanRemote2017,huang2017exploratory}, leading to fresh and unconventional design solutions, provoking new ideas and directions for designers to explore. This could encourage designers to explore beyond their usual style or approach. Participants also suggested that AI agents may be able to adapt to a designer’s style and preferences over time, becoming more aligned with the designer’s creative goals, which may allow AI to offer suggestions that fit the designer’s unique style, expanding the conceptual space in personalised directions.

    \begin{itemize}
        \item \textbf{Design directions:} Design agentic AI systems that are context aware, can learn from user actions and adapt to the designer’s style and preference over time, and produce remote associations, utilising multiple agents with different roles to extend designers' conceptual space.
    \end{itemize}

Our design fiction highlighted the difficulty designers might face in managing inspiration archives and locating information stored in notebooks and portfolios. Designers need an organised way of recalling where the information they want is located. In this case an AI agent could usefully digitise and access the information using text cues described by the user. This will allow users to retrieve the information/stimulus they seek with the agent's support.

\begin{itemize}
    \item \textbf{Design directions:} Develop agentic AI systems that can digitise and organise the user’s external notes and assets. Natural cues such as annotations, sketches, voice descriptions can be used to surface relevant information and spark new creative insights.
\end{itemize}

One of the main reasons designers want to keep creative authority is that they believe AI cannot generate authentic, novel, and useful output. AI produces outputs based on its training data, which can lead to stereotypical and homogenised content~\cite{AgarwalHomogenize2025, Wadinambiarachchi2024}. Moreover, researchers have shown that AI sometimes replicates elements of existing designs too closely, leading to limited diversity in ideas, design fixation~\cite{Wadinambiarachchi2024} and bias~\cite{AgarwalHomogenize2025}. Designers are also concerned that AI will lack insight into user needs, as AI lacks a deep understanding of the emotional and cultural contexts that humans naturally bring to creative work~\cite{Brynjolfsson2017} (see section~\ref{sec:Theme2-1}).

\begin{itemize}
    \item \textbf{Design directions:} Enable the customisability and personalisation of AI agents and train AI models with task-specific knowledge, such as culturally rooted insights, and enable agents to continuously learn from designers’ evolving tastes, and provide ideas that are aligned with the context and designers’ vision.
\end{itemize}

Designers want AI agents to provide multiple “lenses” (view modes with variable fidelities and styles) to broaden their exposure to inspiration; this includes AI-generated ideas tailored to their style, on-demand collation of design components, and providing dynamic perspectives on existing concepts throughout the design process. Previous work on design fixation and ideation has proposed changing the modality, time of exposure, the definition/fidelity of exposure, the distance between the design problem and the stimuli, and the diversity of stimuli~\cite{Vasconcelos2016} to support better ideation. We think it is important to leverage AI capabilities to promote divergent thinking.

\begin{itemize}
    \item \textbf{Design directions:} Create an agentic AI tool that offers designers multiple dynamic lenses with varying fidelity, styles with evolving preferences for designers to broaden inspiration and expand their conceptual space.
\end{itemize}

\subsubsection{Balancing Authority and Agency}
A key finding of our study is that participants want designers to maintain authority and agency over creative decision-making (see section~\ref{sec:Theme2}). They wanted AI agents to take responsibility for mundane, time-consuming, repetitive tasks such as resizing, colour adjustments, and batch processing (see section~\ref{sec:Theme3}). Participants believed that outsourcing these mundane tasks will allow them to focus more on ideation and conceptualisation – the work they want to perform. These results show parallels with the previous findings of Li et al.~\cite{Li2024} and other prior work that has demonstrated that AI tools can boost productivity by handling mundane or technical tasks, freeing up mental energy and time for designers to concentrate on the creative process~\cite{Lu2022, Takaffoli2024, Uusitalo2024}. 

\begin{itemize}
   \item \textbf{Design directions}: Position AI agents as semi-autonomous assistants that can support designers in optimising efficiency and providing technical precision on mundane tasks while leaving core creative decisions in the hands of the designer..
\end{itemize}

\subsubsection{Beyond Text Prompts: Improving Communication with AI Agents.}

Participants felt that text prompting is often not the best way to describe their intent to an AI system, and discussed the need for better UI options (see section~\ref{sec:Theme3}). These findings align with previous HCI research that discusses inconveniences designers face when using prompting to communicate with AI tools~\cite{Li2024,Wadinambiarachchi2024,Takaffoli2024,Uusitalo2024}. Our findings extend existing literature by proposing potential design directions for new interactions. Participants suggested new UI, such as enhanced text prompting, annotations and tagging that allow designers to point at specific parts of the content they refer to. Unsurprisingly, several participants suggested visual sketching as an interface to communicate with AI. And participants felt that voice commands and notes would offer seamless intent sharing. Participants suggested reusing familiar interfaces from existing tools, such as menu-driven, canvas-based, ubiquitous and extended reality interfaces. These designer preferences point to exciting opportunities to extend the design of A-AICSTs.

\begin{itemize}
    \item \textbf{Design directions:} Deploy AI agents capable of interpreting multimodal inputs to understand users’ intent, and provide scaffolding for the users to make better prompts and have clarity in communication.
\end{itemize}

\subsection{A Framework for Orchestrating AI Agents and Human Designers in Design Workflows}
\label{Disc:pt2}

Prior work on human-AI and Mixed-initiative co-creativity have identified factors shaping human-AI collaboration in design. For example, Karimi et al~\cite{karimi2018evaluating} proposed a framework for evaluating creativity in computational co-creativity systems that emphasise on assessing the creative product, process and human-AI interaction to inform the design of self-aware and intentional co-creativity systems. Further, Muller et al.~\cite{ muller2020mixed} proposed a framework that describes collaborative interaction patterns between humans and AI systems based on 11 actions such as learn, ideate, constrain, produce, suggest, etc. and one meta-action–IterativeLoop–, to model turn-taking in mixed-initiative interfaces. Rezwana and Maher introduced Co-Creative Framework for Interaction Design (COFI) ~\cite{rezwana2023designing}, which focuses on modeling interaction dynamics, such as participation style, contribution type, and communication. While these frameworks provide valuable insights into human-AI co creativity, a commonality in all  these frameworks is that they focus on implementation of interactions. 

Our findings identified different roles AI agents could play in design workflows. We analysed designers' views on how agent roles should vary according to the context and task, concluding that the factors that influence an AI agent's role are: the cognitive complexity of the task, the the degree of collaboration when the task is being carried out, the level of creative agency, the distribution of responsibility, and the level of involvement (see section~\ref{sec:Theme1}). Each of these factors can vary. Thus we identified five dimensions that can characterise the form of interaction between human designers and AI agents. We illustrate these dimensions in a conceptual framework extending the stream of literature by providing a novel conceptual framework that is conceptual-oriented rather implementation-oriented. Our framework can be used for defining and orchestrating collaboration in a sociotechnical system comprising AI agents and human designers. 

This framework is intended to support designers and developers in mapping how involvement, responsibility, and creative agency are distributed between human designers and AI agents. It serves as a decision-support tool to guide how humans and AI should interact in collaborative design settings, tailored to the cognitive complexity and the required level of human-AI involvement and degree of shared responsibility.

We present these dimensions in a radial diagram, where each axis radiates from the centre in concentric layers, to show progression along the dimension from low to high intensity.

\subsubsection{The Five Dimensions of the Framework:}

\begin{itemize}
    \item  \texttt{Cognitive Complexity} – This dimension depicts\textit{ the cognitive complexity of the task}. It spans from \textit{mundane}, routine, and repetitive tasks, through \textit{analytical} tasks that involve pattern recognition and insight generation, to \textit{creative} tasks that necessitate original, valuable, and context-appropriate problem-solving.
    \item  \texttt{Degree of Collaboration} – This dimension describes\textit{ the mode of collaboration,} ranging from \textit{individual} (where human or AI works alone), to \textit{coordinated} (where tasks are distributed but interdependent), to \textit{collaborative} (where both humans and agents contribute simultaneously and iteratively).
    \item  \texttt{Creative Agency} – This dimension guides\textit{ decisions on delegation, authorship, and influence over the creative outcome}. It reflects the degree of creative control assigned to human and agents respectively. It scales from low, where the role is minor or supportive, to medium, where both parties contribute creatively, to high, where human or agent takes on substantial generative responsibilities.
     \item  \texttt{Responsibility} – This dimension considers\textit{the responsibility in the decision-making and outcome of the process}. It ranges from \textit{minimal} (tasks with little oversight), to \textit{moderate} (shared responsibility), to \textit{high} (full accountable for decisions and results).
    \item \texttt{Involvement} – This axis explains\textit{engagement level}, from \textit{passive} use (working in a background role), through \textit{neutral} participation, to \textit{active} involvement.
\end{itemize}

\begin{figure*}[htp]
    \centering
\includegraphics[width=0.5\linewidth]{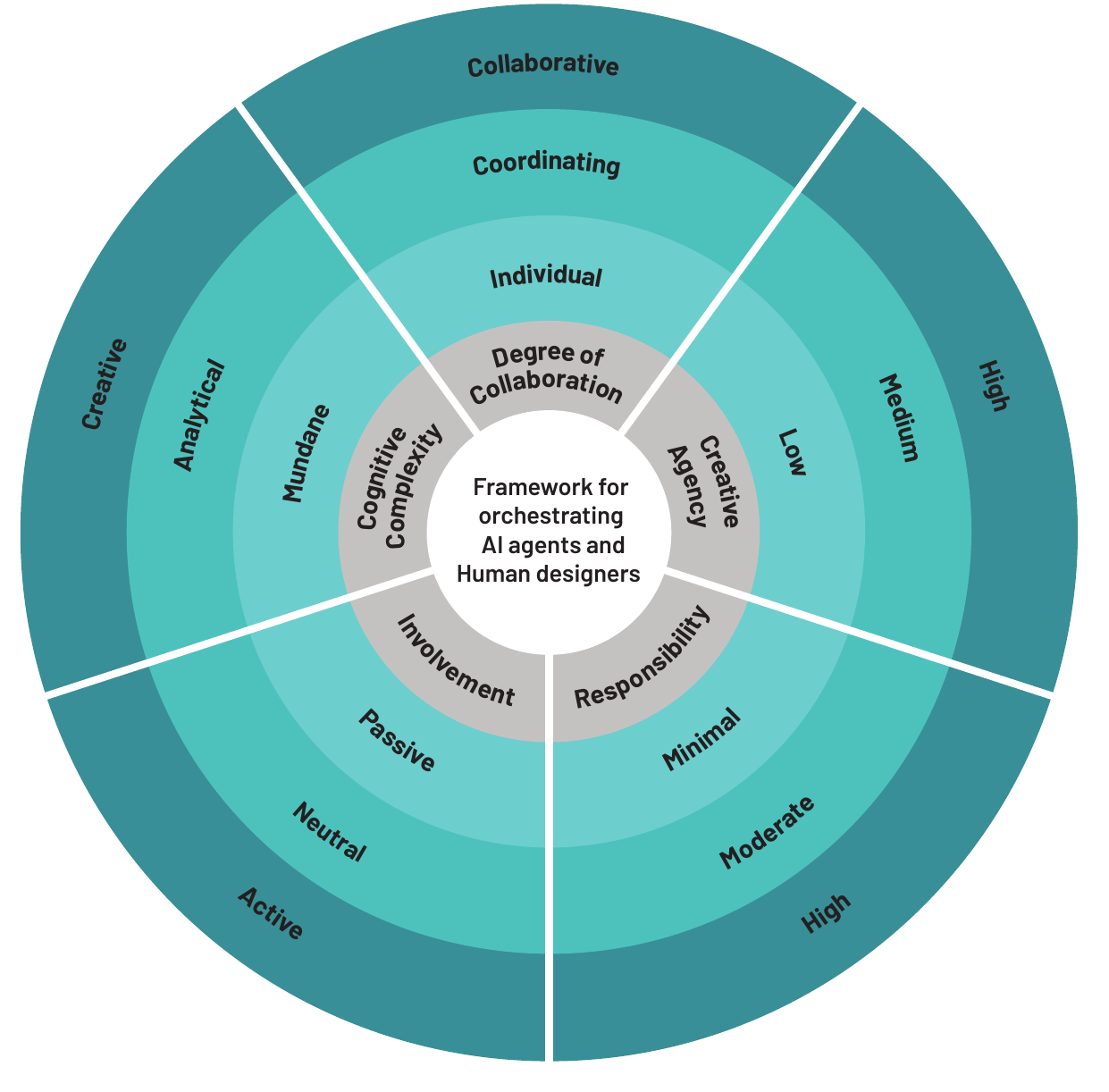}
    \caption{  Conceptual framework for orchestrating AI agents and human designers in design workflows}
    \label{fig:conceptual_framework}
    \Description{This figure shows a Radial diagram illustrating a conceptual framework for orchestrating AI agents and human designers in design workflows. The circle is divided into five wedges, each naming a dimension: Cognitive Complexity, Degree of Collaboration, Creative Agency, Responsibility,and Involvement. Every wedge contains three concentric bands that indicate increasing levels from the center outward. Levels are:
    cognitive complexity-Mundane, Analytical, Creative; Degree of collaboration-Individual, Coordinating, Collaborative; creative agency-Low, Medium, High; Responsibility-Minimal, Moderate, High; Involvement-Passive, Neutral, Active. The diagram communicates that any point in the rings represents a combination of these levels, supporting comparisons of AI and human roles across tasks.
    }
\end{figure*}

\begin{figure*}[htb]
    \centering
\includegraphics[width=0.85\linewidth]{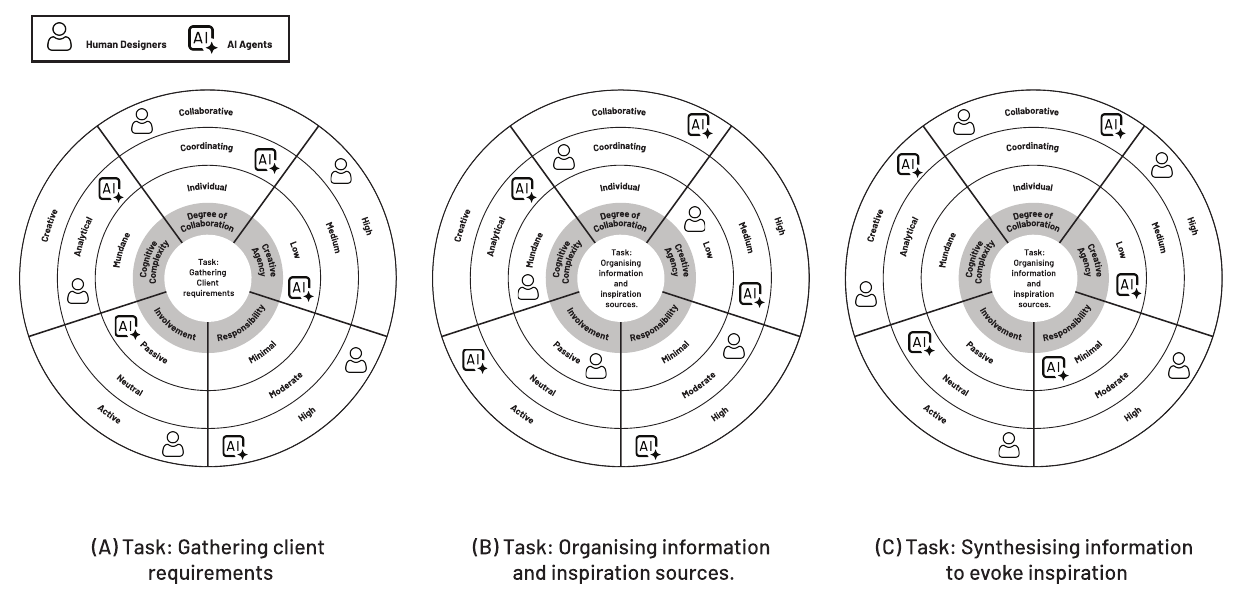}
    \caption{  (A) Gathering client requirements, (B) Organising information and inspiration sources, (C) Synthesising information to evoke inspiration}
    \label{fig:tasks}
    \Description{The figure shows three side-by-side radial diagrams that apply the framework from Figure 8 to specific design tasks: (A) gathering client requirements, (B) organising information and inspiration sources, and (C) synthesising information to evoke inspiration. A legend distinguishes icons, with a person representing human designers and a square badge labelled AI representing AI agents. Icons are positioned at different levels of cognitive complexity, collaboration, agency, responsibility, and involvement to indicate the roles played by AI agents and human designers. Detailed explanations of each task and role allocation are provided in the main text.}
\end{figure*}

We illustrate how this framework can be used by analysing three examples of collaboration between human designers and AI agents, based on design tasks we discussed in our design fiction. Both human and agent are mapped onto the framework.

\subsubsection{Example 1: Gathering Client Requirements.} Figure~\ref{fig:tasks}-(A), In our example, designers are performing an analytical activity to understand what the client wants, by asking different questions, sharing ideas, and capturing the overall goal of the project. Our participants envisioned that the AI agent could be contextually aware, listen to the conversations, and generate meeting summaries. This is a collaborative effort between the designers and the AI agent, which can coordinate the activity. The creative agency of the designer is high, while the agent's creative agency is low. The overall responsibility of the task is with the designer, but the agent may also have high responsibility, as AI needs to capture the conversations correctly and devise plans. The AI agent may get involved passively while the designer performs an active role. 

\subsubsection{Example 2: Organising Information and Inspiration sources.} Figure~\ref{fig:tasks}-(B), In an activity where designers gather inspiration sources and conduct precedence studies, our participants suggested that AI agents can act as assistants, performing routine tasks. If we map these tasks into the dimensions of the framework, we see designers perform a mundane activity of organising their collected information. Here, AI agents could be contextually aware and analytical and cluster the information gathered based on the designers’ usual patterns. The designer can coordinate, while AI agents can play collaborative roles while performing this task. The creative agency of the designer is low, and the agent’s creative agency should be medium as it needs to decide how the categorisation should be done and when to involve the designer. While the responsibility of the AI agent is high, the designer has a moderate responsibility, ensuring all his or her information is appropriately organised. The agent's involvement is active, while the designer can play a passive role. 

\subsubsection{Example 3: Synthesising Information to Evoke Inspiration.}Figure~\ref{fig:tasks}-(C), During the synthesis phase, designers combine and reshape gathered information to spark new ideas. In this phase an AI agent can act as a reframer, providing alternative options, assisting designers to recall what they may have missed while monitoring what they were doing, and supporting designers to expand diverse trajectories of ideas. AI reframers could collaborate with AI coordinator agents to resurface grouped summaries, mood boards, or simple visual maps that are easy to search. This endeavour could be collaborative between human designers and AI agents, with both performing a creative task. The creative agency of humans could be higher, whilst AI agents’ creative agency could be low. The human designer would play an active role, while the involvement of the AI agent can be neutral, adopting both passive and active participation depending on the task.

\subsection{Reflections on Flip-Flap Story Cards Method}
 \label{Disc:pt3}
 
We introduced ``Flip-Flap story cards" in our design fiction study to enhance interactivity and elicit richer insights than traditional interviews or surveys. Participants found this method engaging. The three-sided card design enabled us to deliver the storyline, a question, and a potential solution on the same card, which allowed participants to empathise with characters and ideate based on their own experiences. By hiding the solution on the flap until after participants responded, we encourage participants' free expression before letting them see the potential solution. This way, we were able to strike a good balance between creativity and critical reflection, while making the activity playful, like opening a gift or a Kinder Surprise, despite the preset questions.

To make this method easier for others to adapt, we share practical lessons from our study. First, the storyline should be open enough to encourage diverse interpretations, and the questions need to prompt reflection rather than narrow responses. Second, when developing potential solutions, we incorporated literary techniques such as metaphor and hyperbole to encourage participants’ imagination. For example, in card 20 when Idy locked Dave's access to the tool, participants reflected on how disruptive such an event would be in their own everyday workflow. Third, we learned that conducting pilots in advance is crucial for checking the clarity of the story frames, as well as determining where to place questions and potential solutions within the storyline. For analysis, labelling each response based on the card — whether it involves reading the story, providing an answer, or offering a critique — helps to understand and interpret the meaning of responses more clearly and report the findings systematically. 
In the future, we are interested in exploring the ``Flip-Flap story cards" approach in research contexts beyond speculative design to see whether it could enhance the data collection process.

\subsection{Limitations and Future Directions}
 \label{Disc:pt4}
Our study has some limitations. First, AI design tools are rapidly growing in number and kind, and users are still adapting them to their work. While this timeliness is a strength of our study, it may have reduced the ability of participants to draw on their lived experiences. Second, the representativeness of our participant sample is limited as they were all drawn from one country and were working in large enterprises.

Voluntary participation may have introduce self-selection bias. Finally, we suggest but do not test solutions to the problems discovered; thus, future studies will be needed to verify their effectiveness and practicality.
 
In future research, we aim to validate and test our framework, explore selected design directions to understand the feasibility of implementation, identify areas for further research, and identify the effects of A-AICSTs on the design process. Ongoing research is needed to explore our suggestions for incorporating agentic AI to support creativity in design workflows.

\section{Conclusion}

In this study, we identified the needs and expectations of designers regarding the potential integration of agentic AI into their workflows, through a series of story card elicitation activities set in an imagined future design workspace. We allowed participants to imagine how AI could support designers’ tasks. This speculative approach aimed to provide a "tool for thinking"~\cite{Lallemand2024} to enable participants to creatively extrapolate from current technological and social trends to imagine a future scenario where AI plays a significant role in finding inspiration in design workflows. Our study contributes empirical data on when, where and how designers want support from AI agents to improve their workflow. Moreover, we introduce a new method – using a deck of three-sided Flip-Flap story cards as a tool for thinking within design fiction.

Our work indicates that striking the right balance of authority and agency in creative decision making is crucial in designing agentic AI tools for design work. Therefore, we suggest that developers and vendors of agentic AI systems for designers reflect on whether automation is essential for a given design task and how responsibility, creative agency, and involvement should be orchestrated and distributed between users and agents. Understanding where designers want and do not want AI to intervene may enable vendors of AICSTs to gain a competitive advantage by focusing on specific design tasks rather than seeking to disrupt the entire design workflow. AI agents should support designers in ideation and ease their mundane tasks, but other tasks should be regarded as sacrosanct for designers to do themselves.

\begin{acks}
This research is supported by Melbourne Research Scholarship, the Diane Lemaire Scholarship and the Rowden White Scholarship offered by the University of Melbourne. We would like to express our sincere gratitude to all the participants for their support in this study.
\end{acks}

\bibliographystyle{ACM-Reference-Format}
\bibliography{bibliography.bib}
\newpage
\appendix

\section{Worksheet}
\label{Appendix_A}

\begin{figure}[H]
    \centering
    \includegraphics[width=1\linewidth]{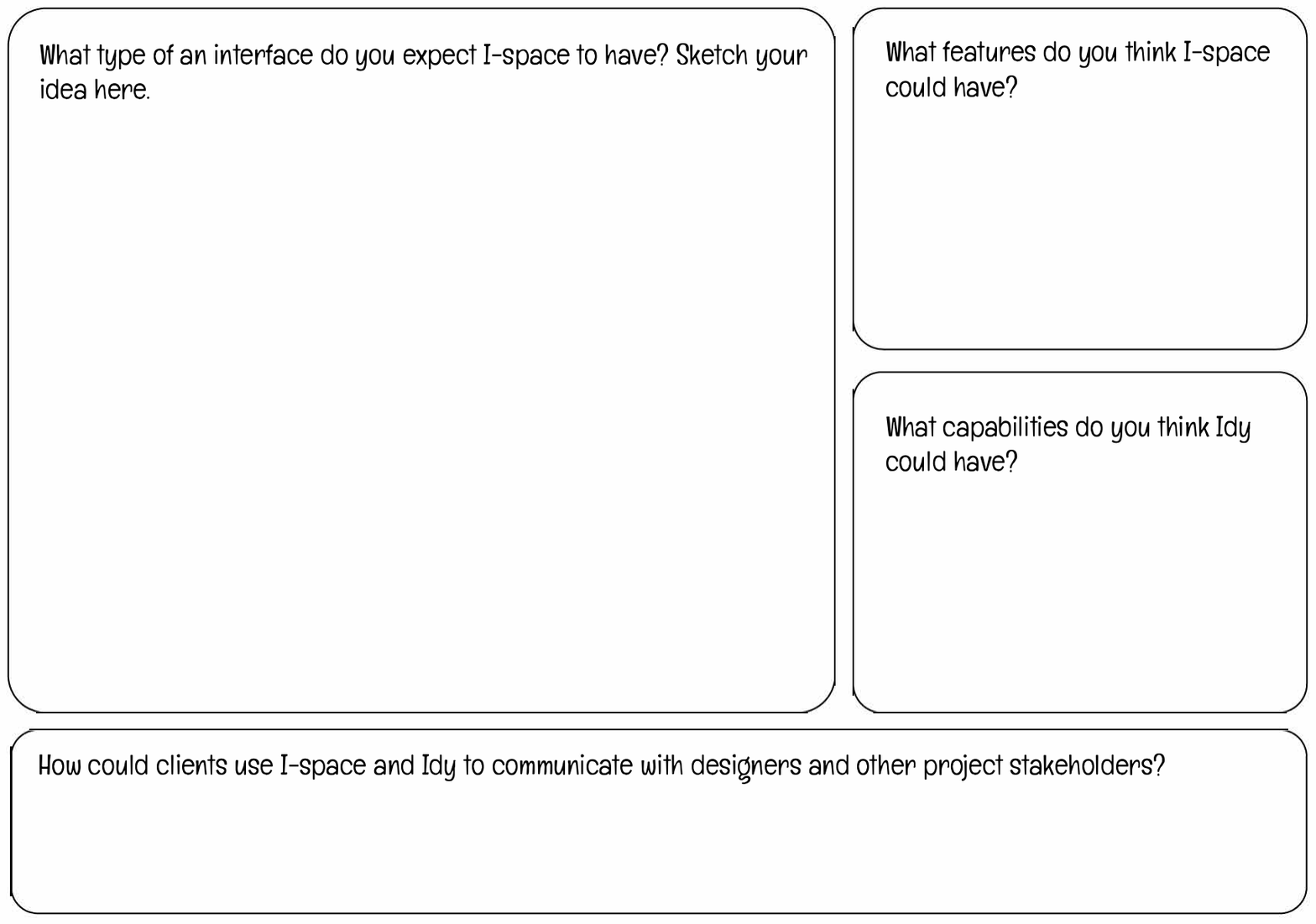}
    \caption*{ Fig. A:  Worksheet provided to participants to complete after the story elicitation task.}
    \Description{The worksheet is divided into four sections, each prompting participants to contribute ideas. The largest box (top left) asks them to sketch the type of interface they expect I-space to have. The top right box prompts for features that I-space could include, and the middle right box prompts for possible capabilities of Idy. Finally, a wide box at the bottom asks how clients could use I-space and Idy to communicate with designers and other project stakeholders. Participants were invited to sketch or write their responses in these spaces.}
\end{figure}

\end{document}